\documentclass[fleqn,usenatbib]{mnras}\usepackage[utf8]{inputenc}
\usepackage[english]{babel}
\usepackage{subfig}
\usepackage{hyperref}

\usepackage{graphicx}
\usepackage{epstopdf}

\usepackage{booktabs}
\usepackage{titlesec}
 
\newcommand\tnumberofnewRBALs{31 }

\newcommand{\angstrom}{\mbox{\normalfont\AA}}

\titleformat{\paragraph}
{\normalfont\normalsize\bfseries}{\theparagraph}{1em}{}
\titlespacing*{\paragraph}
{0pt}{3.25ex plus 1ex minus .2ex}{1.5ex plus .2ex}

\title[new RSBALs from similarity search]{Redshifted broad absorption line quasars found via machine-learned spectral similarity}
\author[I. Reis et al.]{
Itamar Reis$^{1}$\thanks{E-mail: itamarreis@mail.tau.ac.il},
Dovi Poznanski$^{1}$
and Patrick B. Hall$^{2}$
\\
$^{1}$School of Physics and Astronomy, Tel-Aviv University, Tel-Aviv, 69978, Israel\\
$^{2}$Department of Physics and Astronomy, York University, Toronto, ON M3J 1P3, Canada\\
}

\date{Accepted XXX. Received YYY; in original form ZZZ}

\pubyear{2018}

\begin{document}
\label{firstpage}
\pagerange{\pageref{firstpage}--\pageref{lastpage}}

\maketitle

\begin{abstract}
We report the discovery of \tnumberofnewRBALs  new  redshifted broad absorption line quasars (RSBALs)  from the Sloan Digital Sky Survey (SDSS). The number of previously known such objects is 19.  The identification of the new objects was enabled by calculating similarities between  quasar spectra in the SDSS. Using these similarities we look for the objects that are similar to the ones in the original sample, visually inspecting only hundreds, out of over 160,000 spectra considered. We compare the performance of several similarity measures, as well as different methods of employing them, in finding the RSBALs. We find that decision tree based similarities recover the most objects,  and that an ensemble of methods performs better than any single one. As the similarities  are not tailored for the specific problem of finding RSBALs, they could be used for searching for other types of quasars. The similarities and the code for their calculation are available online.
\end{abstract}

\begin{keywords}
methods: data analysis -- methods: machine learning -- (galaxies:) quasars: general -- (galaxies:) quasars: absorption lines -- techniques: spectroscopic
\end{keywords}

\section{Introduction}
There are two topics we discuss in this paper. The first is the detection of new quasars with redshifted broad absorption lines  (RSBALs) from the 14th data release (DR14) of the  Sloan Digital Sky Survey \citep[SDSS,][]{blanton17,abolfathi17}. The second topic is the data mining techniques that enabled the new RSBALs detection in a semi-automated fashion. The introduction is divided accordingly into two parts.  In the first part we discuss the basic properties of RSBALs and broad absorption line (BAL) quasars in general. In the second part we review  approaches for  retrieval of peculiar  objects from large spectroscopic datasets. Peculiar objects in a dataset are, in this context, objects that are not well described by the model we have for the data.

\subsection{Redshifted broad absorption line quasars}
Quasars with broad absorption lines (BALs) are  relatively common, consisting of about 10\% of SDSS quasars \citep[though the fraction changes for non-optically selected quasars and according to][the intrinsic fraction can be as high as 40\%]{allen11}.  BALs are believed to be a consequence of outflows from Active Galactic Nuclei (AGN) accretion disk \citep{murray95}, and as such  are normally blueshifted. 

BAL quasars could be divided into 3 subgroups, based on the transitions in which  absorption appears. The main subgroup is high ionization BAL quasars (HiBALs) in which absorption only appears in transitions from high ionization elements. The second is low ionization BAL quasars (LoBALs) in which both high and low ionization BALs appear. The third is FeLoBAL in which both high and low ionization BALs appear, as well as BALs from   $\mathrm{Fe} \ \mathrm{II}$ and $\mathrm{Fe} \ \mathrm{III}$ transitions.  Almost all of optically selected BAL quasars are HiBALs, 1--3\% are LoBALs, and 0.3\% are FeLoBALs \citep{trump06}. Infrared selected BAL quasars have a much larger fraction of LoBALs \citep{urrutia09}. 

There are two competing scenarios  that can explain why  BALs are detected in only a fraction of quasars. The first suggests that this is a line of sight effect, i.e., BAL quasars are normal quasars for which our line of sight goes through the outflows \citep[][]{weymann91}. The second explanation suggests that BAL quasars are young quasars that will evolve to be regular quasars \citep[e.g.,][]{lazarova12}. In this scenario the BALs are the result of gas and dust surrounding the quasar at an early stage in its evolution, before being expelled. A popular evolutionary scenario connects Ultra Luminous Infra Red Galaxies (ULIRGs) to quasars \citep{sanders88}. In this context BAL quasars are an intermediate stage between ULIRGS and regular quasars. This is supported by the finding of LoBALs residing in ULIRGs \citep[e.g.][]{canalizo02}.

Recently, 19 quasars that show redshifted BALs were discovered by \citet{hall13}. The source of the RSBALs is still unknown. These quasars come from about 12,000 BAL quasars in Data release 9 Quasar Catalog \citep[DRQ9;][]{paris12}, suggesting that redshifted absorption appears in about 1 in 1000 BAL quasars. The LoBAL and RSBAL phenomena are related by the fact that 12--14 of the 19 known RSBAL quasars are LoBAL quasars, compared to a much lower LoBAL fraction in the SDSS sample.  

 \citet{hall13} suggested the following possible explanations: (i) Rotationally dominated outflow. These are outflows that have both radial and rotational velocity components. For such outflows it is possible to get outward velocity along some lines of sight. (ii) Infall of material along the line of sight. The redshifted velocities observed for the RSBALs are up to $\sim$ 10,000 km/s; such large infall velocities could only be produced in the vicinity of the super massive black hole (SMBH). (iii) Binary quasars. In this scenario the RSBALs are outflows from the quasar that is closer to us. The outflows we see as RSBALs are moving away from us in the direction of the other quasar and absorbing the emission coming from it.  \citet{zhang17} presented \textit{Chandra}  X-ray follow up for 7 of these RSBALs.  All the objects are detected to be X-ray weak. This supports the rotationally dominated outflow model for RSBALs, argues against the infall model, and rules out the binary quasar model for these objects.


\subsection{Object retrieval in astronomical datasets}

The 19  original sample RSBALs were found by visually inspecting over 100,000 quasar spectra. This brute force method of finding objects by visually inspecting the entire dataset is of course very (human) time consuming, and not scalable to next generation surveys, but evidently still successfully being used. In recent SDSS quasar catalogs, quasars with BALs or  Damped Lyman-$\alpha$ systems (DLAs) are also flagged  by visual inspection \citep{paris17}. 

How can we automatically  retrieve such objects of special interest out of large datasets? In astronomy a common approach to do so is a query based on fit parameters. This can work well for a large fraction of the data, but, as fitting a model requires making assumptions about the data, the model usually cannot account for all the objects nor all of the features. The objects in question in this work, RSBALs, are an example, as the SDSS fit for quasar spectra does not include absorption lines. In general, models are not available for rare or unexpected objects.  In addition, a model will usually  not cover the entire range of parameters  available in the dataset (a parameter could be temperature for stellar spectra, for example).  This leaves a fraction of the objects, even if well understood, not well fitted, and impossible to query using the database.


 There are a number of approaches for retrieval of objects of a specific type.  The first of these is developing an algorithm that detects objects showing specific features. This can be thought of as developing a model specifically for the objects in question. For example, \citet{el-badry18} searched the Apache Point Observatory Galactic Evolution Experiment \citep[APOGEE, ][]{majewski16} dataset for binary stars by fitting a sum of multiple stellar templates to the spectra. This is in contrast to the single template fitted by the APOGEE pipeline. Another approach is using supervised Machine Learning (ML) algorithms. Here we do not need a physical model of the objects of interest. Instead a sample of known examples is used to train a classifier to recognize these objects. The classifier is then applied to the rest of the data in order to find additional objects of the same type. This approach was applied for example in \citet{parks18}, in which a deep Convolutional Neural Network was trained to detect DLAs. \citet{metcalf18} is an example for a comparison of all of the above methods, including visual inspection, in the context of finding strong gravitational lenses in images. It is interesting to note that they find  automated tools (such as Convolutional Neural Network and Support Vector Machine) to perform better than visual inspection in some aspects.


Another method, which we apply in this work,  is similarity based object retrieval (also called nearest neighbors search). With this method one calculates  a pair-wise similarity  between the objects in the dataset. Starting with a sample of objects of a given type, the similarities allow us to enlarge the sample by finding similar objects  in the rest of the data.  The similarity based method has the advantage that it is 'cheap', meaning the similarity matrix only needs to be calculated once, and could then be applied to all object types (in contrast to the need to develop a different model, or train a new classifier, for every object type). In addition, we can start looking for similar objects given a single example, while with training a detector one usually needs a significant amount of examples in order to avoid issues that arise from imbalanced training sets. The different methods could also be complementary, for example one could first do a similarity search to build a large sample, and follow that with training a detector. 

We note that similarity based object retrieval is sometimes performed in combination with dimensionality reduction tools such as self organizing maps \citep[SOM,][]{kohonen82} and t-distributed stochastic neighbor embedding \citep[t-SNE,][]{maaten08}. In such applications a dimensionality reduction tool is applied to the data, and the similarity is calculated in the lower dimensional space. Examples for applications of these techniques in astronomy are \citet{meusinger12}, which created a SOM of SDSS quasars, and detected additional examples of various types of unusual quasars. In \citet{reis18}  an application of t-SNE to APOGEE infrared stellar spectra enabled detection of previously unknown B-type emission line stars. For our current goal of detecting RSBALs, we find that dimensionality reduction is not an effective technique (we tried  using t-SNE for this purpose). We suggest that this is because the features we are interested in here, i.e, the redshifted broad absorption line, are not dominant features in  quasars spectra (compared to  the continuum shape and emission lines properties).   Dimensionality reduction can discard information about such less dominant features.

\section{Similarity based object retrieval}
In this section we discuss different measures for similarity between two spectra, as well as different approaches for using these similarities for object retrieval.

\subsection{Similarity measures}

The first step in similarity based data mining is  deciding  on a way to measure similarity between objects in the data. The most natural and commonly used similarity measure is euclidian distance (or $\chi^2$,  taking uncertainties into account). Although euclidian distance can be a competitive approach, more sophisticated methods  produce better results for various datasets and applications. As we show below, for the application of detecting RSBALs, measuring (dis)similarity by euclidian distance produces significantly inferior results to all other methods we tried. One thing that makes  measuring similarities between two spectra challenging   is the fact that the information in spectral data is not uniformly distributed. For example, many wavelengths contain information only about the continuum, and when giving all of them the same weight (as in euclidian distance), this fact can mask out the information about emission and absorption lines that exists at a relatively small number of wavelengths. 

It is known that different similarity measures perform well on different data (for example, one can  evaluate   performance of  similarity measures  by the   accuracy of 1st  nearest neighbor classification; that is, given an object, check whether  its 1st nearest neighbor is of the same class). In the data mining literature there are many suggestions for similarity measures for time series, as such data are common in many fields. There is little, if any, discussion regarding spectra. One can classify similarity measures according to their invariances. For example, measuring similarity using cross-correlation is invariant to translations. See  \citet{batista14}  for a review of similarity measures in this context.

In this work we test a number of similarity measures which are described below. Two similarity measures  are based on decision trees; we propose that such measures are well suited for spectral data. The first, Random Forest similarity, have been shown to perform well on spectra in \citet{baron17a} and \citet{reis18}. The second is called Extremely Randomized Trees similarity.

\paragraph*{Random Forest similarity}

 {\fontfamily{cmtt}\selectfont Random Forest} (RF) is a widely used algorithm for classification and regression, see \citet{breiman84,breiman01}.   {\fontfamily{cmtt}\selectfont RF similarity} is an unsupervised application of RF that allows calculating similarities. It was proposed in \citet{breiman03} and \citet{shi06}, and used for astronomical spectra by \citet{baron17a} and \citet{reis18}.  The similarity is calculated by the following procedure. First, synthetic data are created with the same marginal distributions as the original data in every feature, but stripped of the correlation between different features (the features in our application are the flux values at each wavelength of the spectra, as described below). Having two types of objects, one real and one synthetic, an  RF classifier is trained to separate between the two. In the process of separating the synthetic objects with un-correlated features from the real ones, the RF learns to recognize correlations in the spectra of real objects. The  RF is composed of a large number of decision trees and each decision tree is composed of a large number of nodes.  Each tree is trained to separate real and synthetic objects using a subset of the data (the 'Random' in 'Random Forest' is referring to the randomness in which a subset of the data is selected for each tree, see \citet{breiman84,breiman01} and discussion below). Having a large number of trees, the similarity between two objects (objects in the original dataset, i.e., real objects) is then calculated by counting the number of trees in which the two objects ended up on the same leaf (a leaf being a tree node with no child nodes), and dividing by the number of trees. This is done only for the trees in which both objects are classified as real.  We use the {\fontfamily{cmtt}\selectfont scikit-learn} \footnote{\href{http://scikit-learn.org/stable/}{scikit-learn.org/stable/}} \citep{pedregosa11} implementation of  RF.

\paragraph*{Extremely Randomized Trees similarity}
We use another tree-based similarity measure, with a few differences compared to the measure described above. (i) We use a different type of decision trees, called  {\fontfamily{cmtt}\selectfont Extremely Randomized Trees} \citep[ERT, ][]{geurts06}, in which both the feature (in our case, the flux value at  a single rest wavelength) that is used to  split the data at each node, and the split value are chosen at random. (ii) The input to the algorithm is  the \emph{ranking} of objects according to their  value for a given feature, rather than the  feature value itself. (iii) We calculate the similarity by counting for each pair of objects, in each tree, the number of splits the objects go through together. Note that this method does not require creating synthetic data (as the trees are completely random they do not require labels for a training process). 

Motivations for these changes include: (i)   RF learns the important features for the common object types in the sample, and focuses on these to do the classification. As such it might be less sensitive to features that are not common in the dataset, such as absorption lines in unusual locations. ERT selects a flux value at random at each node, and thus does not give less weight to features in unusual locations. (ii) We use rank instead of feature value in order to avoid giving very high weight to very large feature values. The split value is drawn at random between the minimum and maximum feature values, so that for a feature value distribution with large tails, it is likely that the split will be on a value somewhere on the tail, resulting in very unbalanced splits. Note that a regular RF is in practice using the rank and not the feature value, as in an RF the split value is chosen based on the fraction of objects (from each class) that goes to each of the child nodes. (iii) Calculating the similarity by the depth at which two objects are split, instead of looking at  only the  terminal node  is  motivated by the fact that the more nodes two objects pass through together the more the more similar they are, even if they do not end up on the same leaf.  To build the trees for this similarity matrix we used the   scikit-learn implementation of  {\fontfamily{cmtt}\selectfont isolation forest} \citep[ isolation forest is an outlier detection algorithm, here we use its implementation  only to build the extremely randomized trees]{liu08}. To calculate the similarities we use our own code, which is available on  {\fontfamily{cmtt}\selectfont github} \footnote{\href{https://github.com/ireis/ERT-similarity}{github.com/ireis/ERT-similarity}}.


\paragraph*{Additional similarity measures}

We test a number of additional similarity measures. Note that here we show the calculation of the distances instead of similarities. In the context of this work we do not need a transformation between distance and similarity  as we are only interested in the rank. First are three straightforward distance calculations \footnote{These are available in the metrics.pairwise module of {\fontfamily{cmtt}\selectfont scikit-learn}}:

\noindent {\fontfamily{cmtt}\selectfont Euclidean Distance}
\begin{equation}
\mathrm{D_{euclidian}}\left( C, Q \right) = \sqrt{ \sum_i{ \left(q_i - c_i \right)^2 }} 
\end{equation}
$q_i$ and $c_i$ are the features of the two objects $C$ and $Q$.  

\noindent {\fontfamily{cmtt}\selectfont Correlation coefficient}

\begin{equation}
\mathrm{D_{correlation}}\left( C, Q \right) =  1 - \frac{\left(\textbf{q} - \overline{q}  \right) \cdot  \left(\textbf{c} - \overline{c}  \right) }{\left|\textbf{q} - \overline{q}  \right|   \left|\textbf{c} - \overline{c}  \right|}
\end{equation}

\noindent
 {\fontfamily{cmtt}\selectfont Canberra Distance}

\begin{equation}
\mathrm{D_{canberra}}\left( C, Q \right) =  \sum_i{ \frac{  \left| q_i - c_i \right| }{  \left| q_i  \right| + \left| c_i \right| } }
\end{equation}

Another distance measure we try is {\fontfamily{cmtt}\selectfont Dynamic Time Warp} (DTW), a popular measure for time series \citep[see for example][]{berndt94, ratanamahatana04}. This algorithm finds an optimal, under certain constraints, alignment  between the two objects before calculating the (usually euclidian) distance. The possibility to align the spectra before calculating the distance is expected to help cope with bad redshift determination. In addition, BALs can have different offsets and widths, and thus appear at different wavelengths. In this work we want to detect objects that have BALs as similar, even if the BALs have different properties for the two objects. Note that with DTW, a redshifted BAL is not expected to be considered similar to a blueshifted BAL. This is due to the different  relative position with respect to the emission line.  DTW is not sensitive to local offsets of different spectral features (i.e. emission and absorption lines), but it is sensitive to their order. 

{\fontfamily{cmtt}\selectfont Distance Correlation} \citep{szekely10} is a measure of dependance that returns zero if and only if the two input vectors are independent. This is in contrast to the classical measure of dependance, the Pearson correlation coefficient, which is mainly sensitive to a linear relationship. The method has a maximum value of unity.  In the experiments we performed this measure produces similar but slightly better results than the correlation coefficient; therefore, below we include results using the Distance Correlation measure but not the correlation coefficient.

{\fontfamily{cmtt}\selectfont Cross Correlation} is the sliding inner product of the two objects. It measures similarity as a function of displacement between them. We note that for applications of cross correlation on spectra, the correct thing to do is logarithmically shift the objects (i.e, shift the redshift), instead of linearly displace them. In order to use the fast Fourier transform method, we use regular (linear) cross-correlation. This is a valid approach for small shifts only.  We take the maximum value of the cross correlation as the similarity measure. It is also possible to take only the displacement that gives the maximum cross correlation, align the two objects accordingly, and apply any other similarity measure.

\subsection{Methods of similarity search (given a measure)}
Given a similarity matrix and sample of objects of some specific type, there are various ways to search for more objects of the same type. The obvious and widely used method is looking at the nearest neighbors of the original sample objects. There are, however, other methods. One method that we apply here is ranking all objects in the dataset by the number of original sample objects found in their $k$ nearest neighbors (Alternatively one can rank all objects by the number of times they are included in the $k$ nearest neighbors of the original sample objects).
Intuitively, the difference is whether we are looking for objects that are very similar to any single  object from the target sample, or looking for objects that have possibly \textit{weak}  similarity but to \textit{many} objects in the target sample. We say weak similarity, as one can choose a large $k$. 

Choosing the best option depends of the dataset and the characteristics of the target sample. For example, one can consider the uniformity of the objects one is interested in. For a uniform type of objects, looking for the nearest neighbors of the target sample should perform well. In a non uniform type we can have objects that only share some specific, not necessarily dominant, feature. In this case one can expect ranking the objects by number of 'related' target sample objects to perform better. The reason being that if an object is weakly related to many target sample objects, it is likely to share the common features that the target sample objects share.

We apply the two methods discussed above, and find that they are complementary in our case. That is, with each of them we find objects that we do not find with the other.

\section{Data}

We use quasar spectra from the SDSS DR14 \citep{abolfathi17}.  To select quasars we take objects with  \textit{Class = QSO}  from the \textit{SpecObj} table. This criterion  selects objects that were classified as quasars via spectral fitting. We take only quasars for which the rest frame spectrum contains flux values in the  wavelength range: $1120 \angstrom < \lambda  < 2000 \angstrom$. This range contains the following lines:  $\mathrm{C IV \  \lambda 1549 \angstrom}$  in which all RSBAL quasars are expected to have absorption,   $\mathrm{Ly\alpha \ \lambda 1216 \angstrom + N V \ \lambda 1240 \angstrom}$ complex which is important for verification that we are indeed looking at a quasar at the correct redshift, and $\mathrm{C  III] \ \lambda 1909 \angstrom}$ that is useful for redshift determination in BAL quasars, as there is no absorption expected from this transition.  We use the  SDSS pipeline redshift estimate \citep[\textit{z} from \textit{SpecObj},][]{stoughton02}. We take only quasars with signal to noise ratio (SNR, \textit{snMedian} from \textit{SpecObj}) $ > 1.5$, due to the fact that for very low SNR objects it is difficult to confirm the existence of the RSBALs. This leaves us with 164,798 quasar spectra (out of more than 600,000 is the SDSS DR14). Our preprocessing stage consists of removing flux values marked as bad by the SDSS pipeline (i.e., flux values with inverse variance of 0), normalizing the spectra to have a median of one, shifting the spectra to the rest frame, as well as linearly interpolating  the spectra to the same wavelength grid.
 
Out of the 19 BAL quasars with confirmed redshifted absorption, 14 are included in our sample. We list these objects in Table \ref{tab:original_RSBALs}. 2 of the original sample objects are not included as they have low redshift and their wavelength coverage does not include $\mathrm{C  IV}$, these objects were detected by redshifted $\mathrm{Mg  II \ \lambda 2798}$ absorption. 2 others do have $\mathrm{C  IV}$ coverage, but no $\mathrm{Ly\alpha} \ \lambda 1216 \angstrom$. Another object that is not included in our sample does have the wavelength coverage that we require, but has a wrong SDSS redshift estimation.

\section{Results}

\subsection{Experiments}
\label{sec:exp}
We perform a few experiments in order to compare the different similarity measures. The goal of this section is mainly to illustrate that the choice of similarity measure is important, as well as to understand which measures perform well for the specific problem we are addressing.

\begin{figure}%
    \includegraphics[width=\columnwidth]{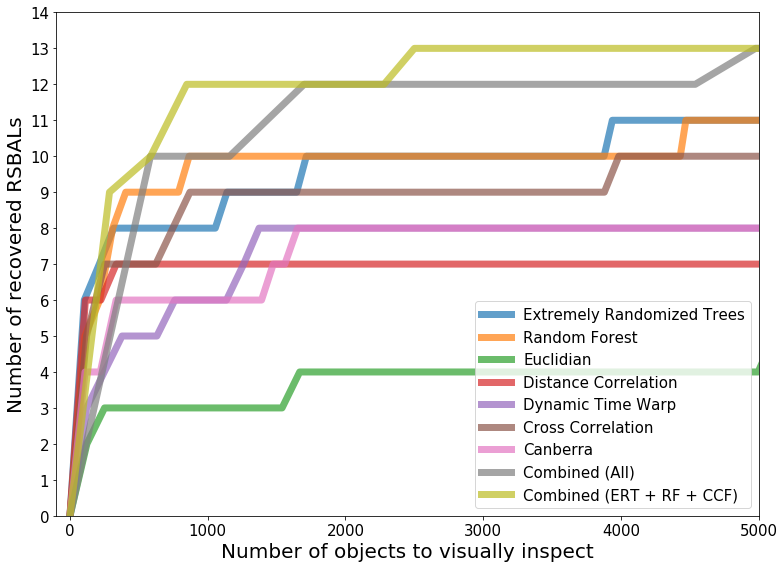}   %
\caption{Comparison of different similarity measures performance in retrieving RSBALs. We count the number of  original sample RSBALs that are recovered by the nearest neighbors of the other original sample objects. We show the number of recovered objects vs. the total number of objects returned by the nearest neighbors search (this corresponds to the number of objects one would need to visually inspect in order to find the RSBALs). }
\label{fig:similarity_tests}
\end{figure}

\paragraph*{Method 1: Nearest neighbors of the original sample}
We measure the performance of the different similarity measures for RSBALs object retrieval with the following experiment.  For each of the 14 original sample objects, we check if it is  recovered by inspecting the nearest neighbors of all the other original RSBALs. We count the total number of recovered objects. This is a similar evaluation metric to the commonly used  {\fontfamily{cmtt}\selectfont precision at top k}.  The results are reported in Figure \ref{fig:similarity_tests}. In this figure we show, for the different similarity measures, the number of recovered objects against the total number of objects returned (in a search for new RSBALs, this would be the number of objects one would need to visually inspect in order to retrieve the RSBALs). The two tree-based similarity measures and the cross correlation measure recover the most objects, while with simple euclidian distance we recover a notably  small number of objects.

The best performing method is the combinations of different  similarity measures. We constructed this combination  by simply concatenating the (unique) nearest neighbors from every measure. Naively, one might find this result surprising, as when combining results from different methods one might expect to obtain an average performance. Instead we get performance that is better than any single method. It is known, however, that in many cases ensemble methods outperform any single model included in the ensemble, and in this context our result seem natural. 

\begin{figure}%
    \includegraphics[width=\columnwidth]{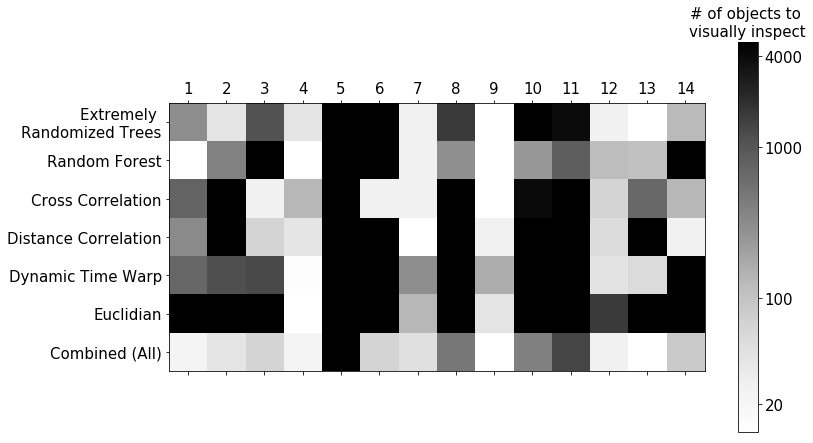}   %
\caption{Comparison between the  different similarity measures we used for object retrieval.  Each entry in this plot shows, for a specific original sample object (horizontal axis) and similarity measure (vertical axis), the number of objects one would need to visually inspect in order to recover the object with the given method. Here we use the nearest neighbors of the original sample objects.}
\label{fig:sim_mes_cmpr_pobject}
\end{figure}

In Figure \ref{fig:sim_mes_cmpr_pobject} we show which objects are found by which similarity measure. There are a few observations one can make by inspecting this figure. (i) One object, J101946.07+051523.6 (\#6 in Figure \ref{fig:sim_mes_cmpr_pobject}), is detected only by the cross correlation similarity measure. This object  has a wrong redshift estimation (We use the SDSS pipeline  estimation of $z = 2.410$, while \citet{hall13} estimate the redshift for this object to be $2.452$). For this reason it is not aligned with the other objects, and thus cannot be  detected as similar by most similarity measures. The cross correlation measure has an invariance for shifts in features, and thus is able to detect this object. (ii) Another object, J094108.92-022944.7 (\#5 in Figure \ref{fig:sim_mes_cmpr_pobject}), is not detected by any similarity measures we apply. J094108.92-022944.7, which is the only original RSBAL to be classified as extended in SDSS imaging, is much redder than the rest of the sample. As all of the similarity measures we applied are sensitive to the continuum, this is most likely the reason it is not recovered. This object also has relatively weak redshifted absorption. (iii) The different methods are complementary. This is true even for the two tree-based measures, each one detects objects that the other does not. (iv) The two tree-based similarity measures combined with the cross correlation measure detect by themselves  all of the objects that are found in total. That is, the other measures do not find any object that is not found by these three measures. Indeed, when we combine the results of only these measures we get better performance than combining all measures (in the sense that we retrieve the same number of objects but with fewer objects to visually inspect, see Figure \ref{fig:similarity_tests}).

\begin{figure}%
    \includegraphics[width=\columnwidth]{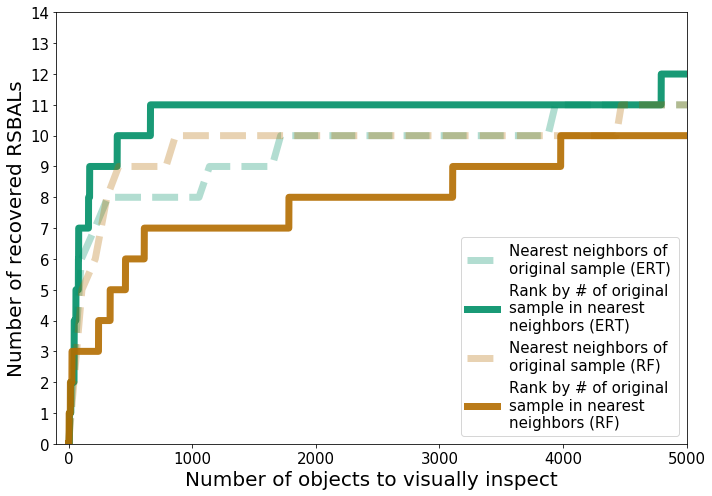}   %
\caption{Comparison of the two methods we tried for applying similarity measures for a nearest neighbor search, for both the RF and ERT similarities.  We show the number of recovered objects vs. the total number of objects returned by the nearest neighbors search (this corresponds to the number of objects one would need to visually inspect in order to find the RSBALs).}
\label{fig:similarity_tests_2nd_method}
\end{figure}

\begin{figure*}%
    \includegraphics[width=\textwidth]{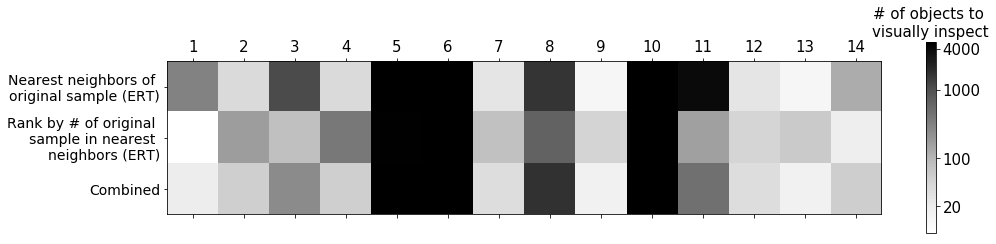}   %
\caption{Comparison between the two different methods we used for object retrieval, using the same similarity matrix (Extremely Randomized Trees).  Each entry in this plot shows, for a specific original sample object (horizontal axis) and object retrieval method (vertical axis), the number of objects one would need to visually inspect in order to recover the object with the given method. We can see that (after deciding on the number of object to visually inspect) some objects would be found only by one of the methods.}
\label{fig:OSNN_vs_NNoOS}
\end{figure*}

\paragraph*{Method 2: Rank all objects by the number of original sample objects in their nearest neighbors}
In Figure \ref{fig:similarity_tests_2nd_method} we show the results of the second method we apply for object retrieval. In this method we rank all objects in the dataset by the number of objects in the original sample that are included in their nearest neighbors, and visually inspect the top ranked objects. We apply this method only to the tree-based measures (the reason for not applying this method to all similarity measures is that this is a more computationally costly method). We can see that, with this method, Extremely Randomized Trees similarity  achieves better results than the Random Forest similarity. Figure \ref{fig:OSNN_vs_NNoOS} shows which objects were found by which method for the  Extremely Randomized Trees similarity. We again see complementary results. Although the difference between the two methods in this experiment is not significant, when looking for new RSBALs we did find several objects with the second method (rank all objects by the number of original sample objects in their nearest neighbors) alone.

\subsection{New BAL quasars with redshifted absorption}

Applying the methods discussed above we find \tnumberofnewRBALs new  BAL quasars with redshifted absorption, and 6 additional RSBAL candidates.  The new RSBALs detections are listed in Table \ref{tab:new_RBALs} (the 6 additional  candidates are listed in Table \ref{tab:RSBALs_CANDS}). In general we  did not keep track of which method found which object, and exactly how many objects were visually inspected in total (we estimate this number to be a few hundreds). Another note is that we found that  inspecting the nearest neighbors of single objects separately to be more efficient (rather than creating one large sample to visually inspect by taking the \# nearest neighbors of each original sample object). This is due to the fact that some original sample objects turned out to have many new RSBALs in their nearest neighbors, while other original sample objects had none. Naturally, we inspected more nearest neighbors of the ones that were productive in finding new RSBALs.  As a final step, the  500 objects with most RSBALs in their ERT nearest neighbors were inspected. For this search we used both the original sample and the newly found objects. We did not find any new RSBALs in this final inspection. Within the first 500 objects this method recovers 42 out of the 48 RSBALs, another RSBAL is recovered in the next few hundred objects, and the rest could not be efficiently recovered using this method  (this test was performed in the same way as the experiments in Section \ref{sec:exp}). We note that one spurious object from \citet{hall13} (SDSS J144424.55+013457.0), out of three in our full sample, was also included in the objects we visually inspected. 

\begin{figure*}
\includegraphics[width=\textwidth]{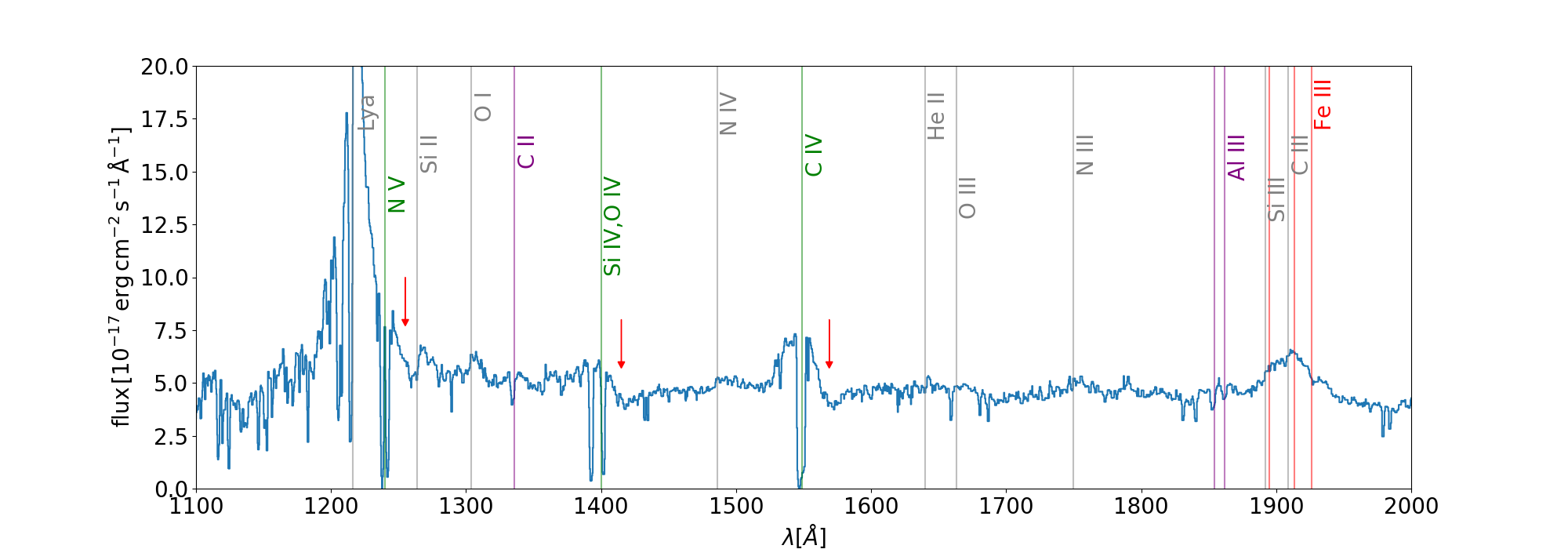}
\caption{SDSS spectrum of SDSSJ12290, the only quasar  in our sample which is not a point source. Red arrows mark the redshifted absorption.}
\label{fig:6081040472247083008}
\end{figure*}
\begin{figure}%
    \includegraphics[width=\columnwidth]{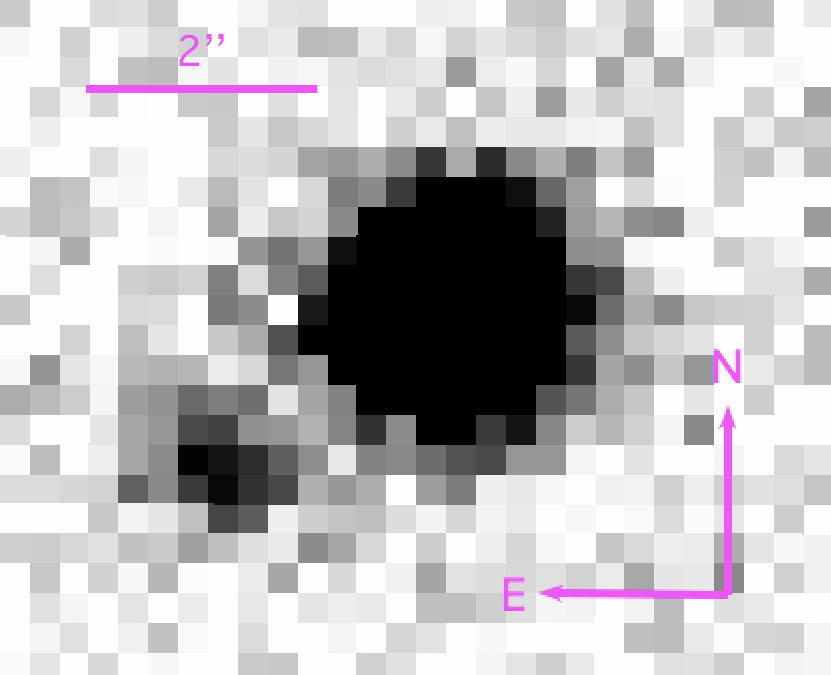}   %
    \caption{DECaLS g-band image of SDSSJ12290 showing. SDSSJ12290 is the only RSBAL in our sample which is not a point source.}%
\label{fig:J1229}%
\end{figure}

In Table  \ref{tab:new_RBALs}  we specify, for each new RSBAL, the transition group (HiBAL, LoBAL or FeLoBAL) in which the redshifted and  blueshifted (if exists) absorptions appears. Between 13 and 20 of the new RSBAL quasars have redshifted LoBALs, and between 9 and 17 of them have blueshifted LoBALs, reinforcing the observation that LoBALs are highly overrepresented among RSBAL quasars. One object of special interest is SDSS J122909.64+093810.1 (SDSSJ12290, \# 14 in Table \ref{tab:new_RBALs}). The spectrum of this object is shown in Figure \ref{fig:6081040472247083008}. This object is not an isolated source, as can be seen in DECaLS \citep{blum16} g-band, Pan-STARRS \citep{chambers16}, and SDSS imaging. Figure \ref{fig:J1229} shows the DECaLS image. This makes SDSSJ1229 a candidate double quasar. This is the only object in our sample which is not an isolated  point source. We also note that this object has a radio  detection in FIRST, with $66.25 \ mJy$ flux. Only one other object in our sample has a radio detection (SDSS J142319.81+223601.2,  \# 26 in Table \ref{tab:new_RBALs}).

\newcounter{magicrownumbers}
\newcommand\rownumber{\stepcounter{magicrownumbers}\arabic{magicrownumbers}}

\begin{table}
\begin{center}

\tiny
\begin{tabular}{l|ccc}

\toprule
index & SDSS name  & Red BAL type  &       Blue BAL type                \\
\midrule
\rownumber &  \href{http://skyserver.sdss.org/dr14/en/tools/explore/summary.aspx?sid=8563799042785517568&apid=}{SDSS J232222.06+232037.2} &           Hi &            Hi \\
 \rownumber &  \href{http://skyserver.sdss.org/dr14/en/tools/explore/summary.aspx?sid=7275653814836764672&apid=}{SDSS J101809.26+282130.0} &           Hi &            Hi \\
 \rownumber &  \href{http://skyserver.sdss.org/dr14/en/tools/explore/summary.aspx?sid=8640426213557379072&apid=}{SDSS J005646.51+282913.5} &           Hi &            Hi \\
 \rownumber &  \href{http://skyserver.sdss.org/dr14/en/tools/explore/summary.aspx?sid=4269645080900378624&apid=}{SDSS J122920.66-012710.0} &           Hi &            Hi \\
 \rownumber &  \href{http://skyserver.sdss.org/dr14/en/tools/explore/summary.aspx?sid=5616024567187218432&apid=}{SDSS J170641.79+345019.7} &           Hi &            Hi \\
 \rownumber &  \href{http://skyserver.sdss.org/dr14/en/tools/explore/summary.aspx?sid=5601531358222458880&apid=}{SDSS J152851.06+344400.7} &           Hi &            Hi \\
 \rownumber &  \href{http://skyserver.sdss.org/dr14/en/tools/explore/summary.aspx?sid=5834616833294966784&apid=}{SDSS J085603.68+292426.7} &           Hi &            Hi \\
 \rownumber &  \href{http://skyserver.sdss.org/dr14/en/tools/explore/summary.aspx?sid=6036112778271367168&apid=}{SDSS J110214.28+110243.1} &           Hi &             - \\
 \rownumber &  \href{http://skyserver.sdss.org/dr14/en/tools/explore/summary.aspx?sid=6064162969750577152&apid=}{SDSS J115652.70+132622.3} &           Hi &             - \\
 \rownumber &  \href{http://skyserver.sdss.org/dr14/en/tools/explore/summary.aspx?sid=6001224450141298688&apid=}{SDSS J100430.91+152100.3} &        Hi/Lo &         Hi/Lo \\
 \rownumber &  \href{http://skyserver.sdss.org/dr14/en/tools/explore/summary.aspx?sid=8037911145573490688&apid=}{SDSS J232044.09+323052.9} &        Hi/Lo &         Hi/Lo \\
 \rownumber &  \href{http://skyserver.sdss.org/dr14/en/tools/explore/summary.aspx?sid=6033782912796565504&apid=}{SDSS J110003.27+080455.1} &        Hi/Lo &         Hi/Lo \\
 \rownumber &  \href{http://skyserver.sdss.org/dr14/en/tools/explore/summary.aspx?sid=5482266779836194816&apid=}{SDSS J091513.00+071237.5} &        Hi/Lo &         Hi/Lo \\
 \rownumber &  \href{http://skyserver.sdss.org/dr14/en/tools/explore/summary.aspx?sid=6081040472247083008&apid=}{SDSS J122909.64+093810.1} &        Hi/Lo &             - \\
 \rownumber &  \href{http://skyserver.sdss.org/dr14/en/tools/explore/summary.aspx?sid=6934474530225102848&apid=}{SDSS J235159.88+094434.2} &        Hi/Lo &          Hi/- \\
 \rownumber &  \href{http://skyserver.sdss.org/dr14/en/tools/explore/summary.aspx?sid=4952897009612726272&apid=}{SDSS J023653.57-062650.1} &        Hi/Lo &            Lo \\
 \rownumber &  \href{http://skyserver.sdss.org/dr14/en/tools/explore/summary.aspx?sid=5486723100966887424&apid=}{SDSS J094922.86+065628.7} &           Hi &       Lo/FeLo \\
 \rownumber &  \href{http://skyserver.sdss.org/dr14/en/tools/explore/summary.aspx?sid=5027327350531923968&apid=}{SDSS J075728.87+252441.1} &           Hi &            Lo \\
 \rownumber &  \href{http://skyserver.sdss.org/dr14/en/tools/explore/summary.aspx?sid=6424406687231221760&apid=}{SDSS J005942.19+105336.0} &           Lo &         Hi/Lo \\
 \rownumber &  \href{http://skyserver.sdss.org/dr14/en/tools/explore/summary.aspx?sid=6900911114613137408&apid=}{SDSS J231610.33+184337.3} &           Lo &    Hi/Lo/FeLo \\
 \rownumber &  \href{http://skyserver.sdss.org/dr14/en/tools/explore/summary.aspx?sid=9264132706702172160&apid=}{SDSS J115858.17+541920.9} &           Lo &         Hi/Lo \\
 \rownumber &  \href{http://skyserver.sdss.org/dr14/en/tools/explore/summary.aspx?sid=8533258732548628480&apid=}{SDSS J221228.68+214021.9} &      Lo/FeLo &            Hi \\
 \rownumber &  \href{http://skyserver.sdss.org/dr14/en/tools/explore/summary.aspx?sid=5688302064087572480&apid=}{SDSS J222447.34+102109.8} &           Lo &            Lo \\
 \rownumber &  \href{http://skyserver.sdss.org/dr14/en/tools/explore/summary.aspx?sid=6010224227726106624&apid=}{SDSS J101922.81+125922.2} &           Lo &            Lo \\
 \rownumber &  \href{http://skyserver.sdss.org/dr14/en/tools/explore/summary.aspx?sid=7302609991692689408&apid=}{SDSS J131852.14+282357.2} &           Lo &            Lo \\
 \rownumber &  \href{http://skyserver.sdss.org/dr14/en/tools/explore/summary.aspx?sid=6771287760827293696&apid=}{SDSS J142319.81+223601.2} &           Lo &            Lo \\
 \rownumber &  \href{http://skyserver.sdss.org/dr14/en/tools/explore/summary.aspx?sid=6155543930303651840&apid=}{SDSS J143829.67+095555.0} &           Lo &            Lo \\
 \rownumber &  \href{http://skyserver.sdss.org/dr14/en/tools/explore/summary.aspx?sid=8260875337711198208&apid=}{SDSS J023959.74+000231.6} &           Lo &            Lo \\
 \rownumber &  \href{http://skyserver.sdss.org/dr14/en/tools/explore/summary.aspx?sid=7118122933352898560&apid=}{SDSS J162853.69+475058.7} &           Lo &             - \\
 \rownumber &  \href{http://skyserver.sdss.org/dr14/en/tools/explore/summary.aspx?sid=6180300809277775872&apid=}{SDSS J152004.60+105634.3} &      Lo/FeLo &     Lo/FeLo/- \\
 \rownumber &  \href{http://skyserver.sdss.org/dr14/en/tools/explore/summary.aspx?sid=7685491828351541248&apid=}{SDSS J131410.90+655100.5} &         FeLo &             - \\
\bottomrule
\end{tabular}
\caption{Newly discovered quasars with redshifted broad absorption lines (RSBALs). These objects were found as neighbors (in a similarity matrix) of the 19 previously known RSBALs from \citet{hall13}.  The Red/Blue BAL type columns refer to the group of transitions in which the Red/Blue BALs appear.  Note that always when there are FeLoBALs there are also LoBALs, and similarly, LoBALs include HiBALs. The origin of the RSBALs is still unknown, one clue is the overrepresentation of LoBALs in quasars that have RSBALs.   The LoBAL fraction in the SDSS sample (in  which these objects were discovered) is a few percent.}
\label{tab:new_RBALs}

\large
\end{center}
\end{table}


\section{Summary}


We present a sample of \tnumberofnewRBALs new  RSBALs, a significant addition to the 19 known ones. These objects were found using a nearest neighbors search starting from the original sample of known RSBALs. Using this technique we were able to find the new objects by visually inspecting only a few hundred  quasars. This is compared to the visual inspection of 100,000 quasars that was required to build the original sample. 

We compared the performance of several different similarity measures in recovering the original sample of RSBALs. We found that the two tree-based similarity measures, Random Forest similarity and Extremely Randomized trees similarity, deliver the best results. This is while a simple euclidian similarity measure  retrieves a particularly low number of objects. The performance of a few other similarity measures we applied fall between the tree-based ones and the euclidian one. The best results  in the experiments we performed were obtained by an ensemble of  different methods. This suggests that even after finding the best performing similarity measures for the specific problem it is still worthwhile to include results from other measures and use an ensemble of methods. In other words, it is better to look as the top ranked objects from each method, than to dig into lower ranked objects of  a single method.

It is important to note that none of the similarity measures we used was  tailored for this specific problem, and as such are directly applicable for searching other types of quasars. We publish the 100 nearest neighbors, from  the ERT  similarity measure,  of all 164,798 objects in our sample. In addition we produce a  {\fontfamily{cmtt}\selectfont Jupyter Notebook} \footnote{\href{https://github.com/ireis/SDSS-quasars-similarity}{github.com/ireis/SDSS-quasars-similarity}} which can be used to retrieve similar objects. Note that for this work we used the redshift from the SDSS pipeline in order to measure similarities in the rest frame of the spectra. We used only objects with rest frame wavelength coverage of $1120 \angstrom > \lambda  > 2000 \angstrom$. In future work we plan to produce an exhaustive similarity matrix for SDSS quasar spectra.

We emphasize that the reason for not using a larger fraction of SDSS quasars in this work is that the similarities measures we used, except cross-correlation and DTW,  require the objects features to be aligned. For this we need all objects in our sample to contain flux values on a pre-defined wavelength grid. This allows us to use objects of only a limited redshift range. In addition, this alignment requirement  compels us to rely on the SDSS pipeline redshift in our pre-processing stage. This will lead to meaningless similarity values for objects with wrong pipeline redshift. The objects with wrong redshift are the ones that are not well described by the pipeline, and as such are of  high interest for investigations such as the one performed in this work. For this reason calculating similarities that are invariant to shifts in features is an important goal for future work.

\section*{Acknowledgements}
IR would like to thank Natalie Lubelchick, Dalya Baron, and Sahar Shahaf for their advice and help. PBH acknowledges support from the Natural Sciences and Engineering Research Council of Canada (NSERC), funding reference number 2017-05983

This research made use of: the NASA Astrophysics Data System Bibliographic Services, scikit-learn \citep[][]{pedregosa11}, SciPy \citep[including pandas and numpy][]{scipy01},  IPython \citep[][]{perez07}, matplotlib \citep[][]{hunter07}, astropy \citep[][]{astropy-collaboration13}, numba \citep{lam15}, and the SIMBAD database \citep[][]{wenger00}.

This work made extensive use of SDSS data. Funding for the Sloan Digital Sky Survey IV has been provided by the Alfred P. Sloan Foundation, the U.S. Department of Energy Office of Science, and the Participating Institutions. SDSS-IV acknowledges
support and resources from the Center for High-Performance Computing at
the University of Utah. The SDSS web site is www.sdss.org.

SDSS-IV is managed by the Astrophysical Research Consortium for the 
Participating Institutions of the SDSS Collaboration including the 
Brazilian Participation Group, the Carnegie Institution for Science, 
Carnegie Mellon University, the Chilean Participation Group, the French Participation Group, Harvard-Smithsonian Center for Astrophysics, 
Instituto de Astrof\'isica de Canarias, The Johns Hopkins University, 
Kavli Institute for the Physics and Mathematics of the Universe (IPMU) / 
University of Tokyo, Lawrence Berkeley National Laboratory, 
Leibniz Institut f\"ur Astrophysik Potsdam (AIP),  
Max-Planck-Institut f\"ur Astronomie (MPIA Heidelberg), 
Max-Planck-Institut f\"ur Astrophysik (MPA Garching), 
Max-Planck-Institut f\"ur Extraterrestrische Physik (MPE), 
National Astronomical Observatories of China, New Mexico State University, 
New York University, University of Notre Dame, 
Observat\'ario Nacional / MCTI, The Ohio State University, 
Pennsylvania State University, Shanghai Astronomical Observatory, 
United Kingdom Participation Group,
Universidad Nacional Aut\'onoma de M\'exico, University of Arizona, 
University of Colorado Boulder, University of Oxford, University of Portsmouth, 
University of Utah, University of Virginia, University of Washington, University of Wisconsin, 
Vanderbilt University, and Yale University.

\bibliography{red_}
\medskip
\bsp	


\appendix
\section{Original RSBALs sample}
In Table \ref{tab:original_RSBALs} we list the RSBALs from \citet{hall13}  that are included in our sample. We used these objects in the search for new RSBALs.
\setcounter{magicrownumbers}{0}

\begin{table}
\begin{center}

\tiny
\begin{tabular}{l|ccc}

\toprule
index & SDSS name  & \citet{hall13} index        \\
\midrule
 \rownumber &  SDSS J014829.81+013015.0 &                     2 \\
 \rownumber &  SDSS J080544.99+264102.9 &                     3 \\
 \rownumber &  SDSS J082818.81+362758.7 &                     4 \\
 \rownumber &  SDSS J083030.26+165444.8 &                     5 \\
 \rownumber &  SDSS J094108.92-022944.7 &                     6 \\
 \rownumber &  SDSS J101946.07+051523.6 &                     7 \\
 \rownumber &  SDSS J114655.05+330750.0 &                     9 \\
 \rownumber &  SDSS J114756.00-025023.4 &                    10 \\
 \rownumber &  SDSS J132333.01+004633.8 &                    11 \\
 \rownumber &  SDSS J143945.28+044409.2 &                    12 \\
 \rownumber &  SDSS J144055.59+315051.7 &                    13 \\
 \rownumber &  SDSS J170953.28+270516.6 &                    15 \\
 \rownumber &  SDSS J172404.42+313539.6 &                    16 \\
 \rownumber &  SDSS J215704.27-002217.7 &                    17 \\
\bottomrule
\end{tabular}
\caption{Objects from the original RSBALs  sample \citep{hall13} that are included in our sample and used in the search for new RSBALs.}
\label{tab:original_RSBALs}

\large
\end{center}
\end{table}

\section{Additional RSBALs candidates}
In Table \ref{tab:RSBALs_CANDS} we list additional objects detected in this work with suspected RSBALs. Similar to the objects in Appendix B of \citet{hall13}, they show evidence for at least one redshifted BAL trough, but cannot be regarded as secure identifications due to reasons such as trough weakness, spectrum signal-to-noise ratio, redshift uncertainty, and spectral complexity.
\setcounter{magicrownumbers}{0}

\begin{table}
\begin{center}

\tiny
\begin{tabular}{l|ccc}

\toprule
index & SDSS name \\
\midrule
 \rownumber &  \href{http://skyserver.sdss.org/dr14/en/tools/explore/summary.aspx?sid=5479895133724844032&apid=}{SDSS J084700.75+063556.6}        \\
 \rownumber &  \href{http://skyserver.sdss.org/dr14/en/tools/explore/summary.aspx?sid=7235140659778134016&apid=}{SDSS J104045.96+221802.6}            \\
 \rownumber &  \href{http://skyserver.sdss.org/dr14/en/tools/explore/summary.aspx?sid=7347722957387440128&apid=}{SDSS J003043.37+300549.8}            \\
 \rownumber &  \href{http://skyserver.sdss.org/dr14/en/tools/explore/summary.aspx?sid=4142301293388472320&apid=}{SDSS J075117.59+513739.2}            \\
 \rownumber &  \href{http://skyserver.sdss.org/dr14/en/tools/explore/summary.aspx?sid=7039375911544659968&apid=}{SDSS J004232.28+302546.5}            \\
 \rownumber &  \href{http://skyserver.sdss.org/dr14/en/tools/explore/summary.aspx?sid=6033782912796565504&apid=}{SDSS J110003.27+080455.1}            \\

\bottomrule
\end{tabular}
\caption{Additional RSBAL candidates found in this work.}
\label{tab:RSBALs_CANDS}

\large
\end{center}
\end{table}

\label{lastpage}

\end{document}